\documentclass[10pt, conference, letterpaper]{IEEEtran}
\IEEEoverridecommandlockouts

\usepackage{cite}
\usepackage{amsmath,amssymb,amsfonts}
\usepackage{algorithmic}
\usepackage{graphicx}
\usepackage{textcomp}
\usepackage{xcolor}
\def\BibTeX{{\rm B\kern-.05em{\sc i\kern-.025em b}\kern-.08em
    T\kern-.1667em\lower.7ex\hbox{E}\kern-.125emX}}

\usepackage{enumitem,kantlipsum}
\usepackage{comment}
\usepackage{bbding}
\usepackage{subcaption}
\usepackage[colorinlistoftodos]{todonotes}
\usepackage{balance}
\usepackage{wrapfig}
\usepackage{lipsum}

\newcommand{\brandName}{CMMA}
\newcommand{\brandNameShort}{CMA}
\newcommand{\brandNameCommon}{C(M)MA}

\begin{document}

\title{Caching-based Multicast Message Authentication in Time-critical Industrial Control Systems
\thanks{This research is supported by the National Research Foundation, Prime Minister’s Office, Singapore under its Campus for Research Excellence and Technological Enterprise (CREATE) programme.}
}

\author{\IEEEauthorblockN{Utku Tefek}
\IEEEauthorblockA{\textit{Advanced Digital Sciences Center}\\
Singapore\\
u.tefek@adsc-create.edu.sg}
 \and
 \IEEEauthorblockN{Ertem Esiner}
\IEEEauthorblockA{\textit{Advanced Digital Sciences Center}\\
Singapore\\
e.esiner@adsc-create.edu.sg}
 \and
 \IEEEauthorblockN{Daisuke Mashima}
\IEEEauthorblockA{\textit{Advanced Digital Sciences Center}\\
Singapore\\
daisuke.m@adsc-create.edu.sg}
 \and

\IEEEauthorblockN{Binbin Chen}
\IEEEauthorblockA{\textit{Singapore University of Technology and Design}\\
Singapore\\
binbin\_chen@sutd.edu.sg}
 \and
 \IEEEauthorblockN{Yih-Chun Hu}
 \IEEEauthorblockA{\textit{University of Illinois at Urbana-Champaign} \\
 Illinois, USA \\
yihchun@illinois.edu}
}

\thispagestyle{empty}

\maketitle

\begin{abstract}
Attacks against industrial control systems (ICSs) often exploit the insufficiency of authentication mechanisms. Verifying whether the received messages are intact and issued by legitimate sources can prevent malicious data/command injection by illegitimate or compromised devices. However, the key challenge is to introduce message authentication for various ICS communication models, including multicast or broadcast, with a messaging rate that can be as high as thousands of messages per second, within very stringent latency constraints. 
For example, certain commands for protection in smart grids must be delivered within 2 milliseconds, ruling out public-key cryptography. This paper proposes two lightweight message authentication schemes, named \brandNameShort\ and its multicast variant \brandName , that perform precomputation and caching to authenticate future messages. 
With minimal precomputation and communication overhead, \brandNameCommon\ eliminates all cryptographic operations for the source after the message is given, and all expensive cryptographic operations for the destinations after the message is received. \brandNameCommon\ considers the urgency profile (or likelihood) of a set of future messages for even faster verification of the most time-critical (or likely) messages. 
We demonstrate the feasibility of \brandNameCommon\ in an ICS setting based on a substation automation system in smart grids.


\end{abstract}


\section{Introduction} \label{sec:intro}
In industrial control systems (ICSs), enabling the devices to verify that the packets originated from their claimed source and have not been modified while in transit is critical for reliable and trustworthy operation. On the other hand, many ICSs require low-latency, and often high-volume messaging, which pose a significant challenge when integrating security mechanisms into the network. According to the IEEE Power and Energy Society guidelines, periodic status updates and automated control for protection in a field substation of a smart grid, e.g., circuit breaker control upon over-current, requires response time to be 1-2 ms, including all the network-related and processing delays, and involves up to 4,000 messages per second~\cite{ieee1646,fpro2019}. Due to this tight requirement, IEC 61850 --- an increasingly adopted standard for substation automation~\cite{IEC61850original,IEC61850} --- utilizes link-layer multicast for sharing emergency event information with as many as hundreds of devices.

In point-to-point communication settings, the straightforward method of appending a message authentication code (MAC) to each packet using a shared secret key allows the destination to perform this verification. While computationally inexpensive, such symmetric cryptography approaches with a group key are not secure for broadcast and multicast communication settings, because any destination in possession of the shared secret key can impersonate the source and inject forged packets. To prevent this attack, pairwise symmetric keys are necessary. With pairwise symmetric keys, however, the source would need to generate a separate MAC for each destination, increasing the computational load linearly with the number of destinations. Other downsides are the complicated key distribution and storage overhead. Hence, one should look for asymmetry between the source and destinations.

Digital signatures are widely used for multicast/broadcast communication, as signing a packet with a private key enables its verification by any entity that possesses the corresponding public key. Indeed, digital signatures are secure for multicast message authentication, and they offer additional properties such as non-repudiation. IEC 62351-6 standard recommends using RSA signatures for IEC 61850 GOOSE message authentication \cite{cleveland2005iec}. However, one common drawback of the digital signatures is expensive computations, such as modular exponentiation (RSA \cite{RSAsignature}), elliptic curve scalar multiplication (ECDSA \cite{johnson2001elliptic}), and cryptographic pairing, which introduce high overhead for signing and verification, especially on resource-constrained devices (e.g., legacy ICS devices).
There are other methods for generating asymmetry between the source and destinations while retaining the computational efficiency of symmetric cryptography. Hash-chains \cite{hauser1997reducing} and the notion of delayed key disclosure \cite{cheung1997efficient,perrig2002tesla,perrig2000efficient,liu2004multilevel,liu2005practical,TESLA_infocom} have been used to authenticate routing updates in routing protocols. 
These methods generate time asymmetry between the source and destinations, hence achieving public verifiability without having to resort to expensive asymmetric cryptography. However, one major drawback of delayed key disclosure is that the destinations cannot verify a message until the corresponding key is disclosed. Clearly, a disclosure delay in the order of a few messaging cycles is not acceptable for time-critical ICSs. 

Besides the constraints, ICS messages have certain domain-related features, such as structured and predictable message content. ICS messages are semantically fragmented into predefined fields, each with a relatively small set of possible values. Some message fields contain predetermined values such as IDs, sequence numbers, and expiry period known to the source long before the message is sent. Other fields may contain measurements that constantly fluctuate around specific values or several binary flags indicating urgent commands/alerts. The measurement values can be largely predictable due to their almost constant base value, while the urgent binary values only yield a limited number of possible outcomes. The limited entropy of ICS messages can be exploited to accelerate message authentication as elaborated in Section \ref{sec:predictability}. 



In this paper, we first evaluate several baseline caching approaches that precompute and store cryptographic evidence --- a piece of information to verify the source and integrity of the message --- for potential future messages. These relatively primitive designs incur significant precomputation and communication overhead in order to minimize the computations after the message is given (post-message). Then, we introduce Caching-based Message Authentication (\brandNameShort) and the multicast variant Caching-based Multicast Message Authentication (\brandName). Given a set of potential future messages, both schemes precompute and cache cryptographic evidence for a set of possible messages, hence building a cache for message authentication. The proposed schemes employ authenticated data structures to mitigate the overhead suffered by the baseline designs. In particular, we use Huffman Hash Tree (HHT) to ensure faster verification of urgent or likely messages over relatively time-tolerant or less-likely messages. \brandNameCommon\ relies on symmetric keys hence does not suffer from the drawbacks of asymmetric cryptography. \brandNameShort\ is based on MACs, and is suitable when the communication is unicast or the number of destinations is small. \brandName\ employs an adaptation of delayed key disclosure. Thus its overhead does not depend on the number of destinations, making it suitable for multicast/broadcast communication, however it requires loose time synchronization. Despite the delayed disclosure of keys, \brandName\ does not suffer from the disclosure delay of such schemes, as discussed in Section~\ref{sec:premmaTESLA}. To our knowledge, we are the first to demonstrate the feasibility and practicality of multicast message authentication using a precomputation cache for low-latency, high-rate ICS communication on resource-constrained devices.


\section{Approach}\label{sec:approach}

\subsection{Design Goals and Threat Model} \label{sec:GOOSEdisc}
In ICSs in general, ensuring message integrity and authenticity is critical for defending against threats. For instance, verifying that the commands/messages have been initiated only by trusted devices and have not been altered by an unauthorized party can thwart malicious command injection and false data injection attacks. 
%
In the power grid system, as is the case with many other ICSs, timely communication among the devices is imperative. In the modernized power grid systems, IEC 61850 GOOSE  (Generic Object Oriented Substation Event), a link-layer multicast, publisher-subscriber communication protocol, is used among a predefined group of devices, e.g., intelligent electronic devices (IEDs) and programmable logic controllers (PLCs), for regular status updates and urgent control communication. The status updates are announced both regularly and in an on-demand manner whenever the status or measurement of the power grid device is updated, and messages for propagating events such as over current and automated protection control (e.g., opening circuit breakers) require very short latency (1-2 ms)~\cite{ieee1646,li2011performance}.
Besides the end-to-end latency, throughput requirement also poses a challenge. For instance, IEC 61850 SV (Sampled Value) protocol has almost the identical message structure and communication model, but it is sent with a constant and much higher rate (e.g., 4,000 messages/sec)~\cite{IEC61850original, IEC61850}. 



We consider typical remote attackers for false data injection attacks~\cite{roomi2020}, where attackers have a footprint in the system (e.g., via compromised VPN~\cite{case2016analysis} and malware~\cite{CrashOverride}). Such remote attackers could inject arbitrary packets and observe network traffic. However, they are unable to manipulate the configuration of devices (e.g., installation of malicious software/firmware) since the configuration of ICS devices is typically done through local, serial connections. Against this threat model, we aim to design a message authentication mechanism that incurs minimal overhead even for multicast traffic and is also verifier-efficient. Such a defense mechanism is to be deployed on ICS devices or Bump-in-the-wire (BITW) devices in front of them. Note that BITW devices are not addressable and thus not accessible to remote attackers in our scope.

\subsection{Entropy of ICS Messages} \label{sec:predictability}
At the high level, our approach to reducing the latency in authenticating ICS messages involves precomputation and caching of the cryptographic evidence for potential future messages. Such a strategy could be feasible when the timing and content of ICS messages are, to some extent, predictable. While we target ICSs in general, in this section, we discuss the entropy of ICS messages by using IEC 61850 GOOSE as a concrete example.

\begin{figure}[ht]
\begin{center}
\includegraphics[width=0.7\linewidth]{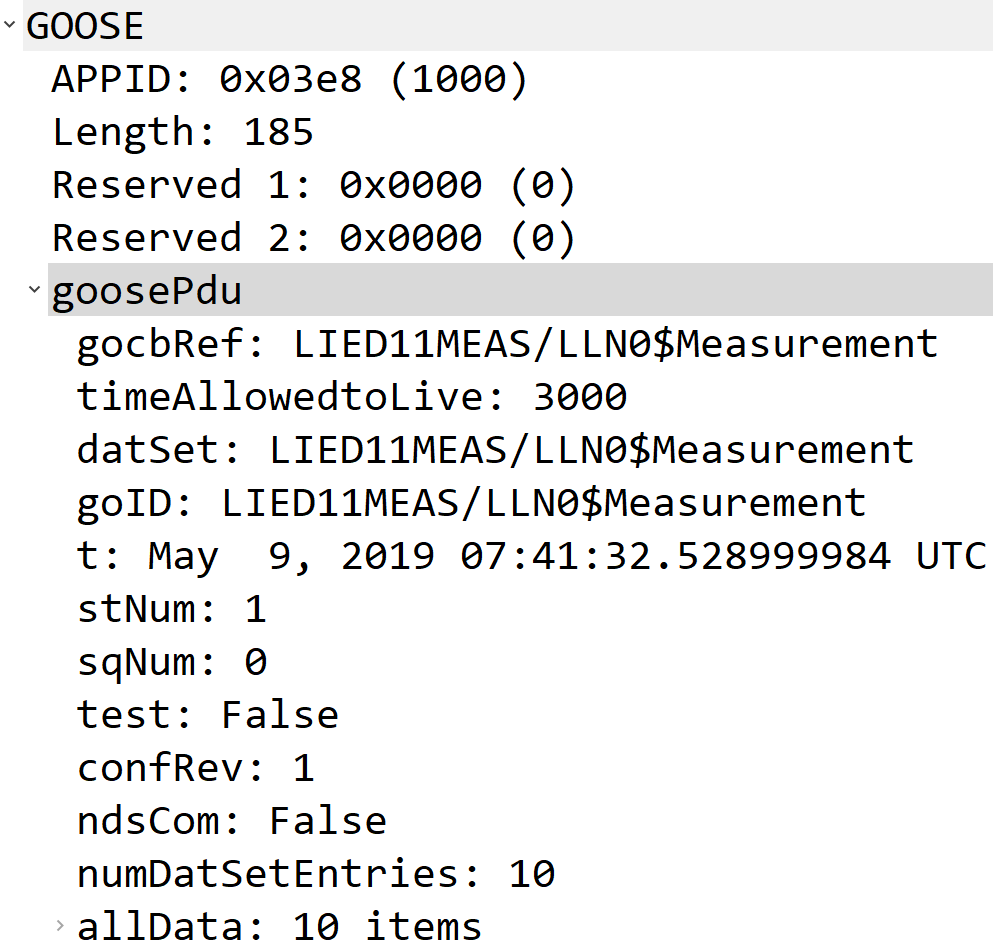}
\caption{The contents of the GOOSE Protocol Data Unit.}
\label{fig:GOOSEpacket}
\end{center}
\vspace{-15pt}
\end{figure}

As shown in Fig.~\ref{fig:GOOSEpacket}, a GOOSE Protocol Data Unit (PDU) consists of a GOOSE control block reference ({\it gocbRef}), a two-byte long {\it timeAllowedtoLive} field specifying the lifetime of the message, an identifier of the dataset included ({\it datSet}), a GOOSE ID ({\it goID}), an 8-byte long timestamp ({\it t}), a status number ({\it stNum}), a sequence number ({\it sqNum}) which is incremented by one or rolled over to zero upon each packet transmission, a test bit {\it test}, configuration revision ({\it confRev}) and needs commissioning ({\it ndsCom}) flags, and the number of user-defined data entries ({\it numDatSetEntries}) \cite{GOOSEMessageStructure,biswas2019synthesized}. The last portion of the GOOSE packet is the {\it allData} field, which stores device/alarm status and measurements.

The prediction of GOOSE PDU fields is trivial except for the timestamp {\it t} and user-defined {\it allData} field. For example, either the status or sequence number is incremented by one in each packet, {\it gocbRef} and {\it datSet} fields are known from the IED's configuration file (called IEC 61850 SCL file), 
and {\it confRev, ndsCom, test, timeAllowedtoLive} values are static for a given system and source-destination(s) setting.
\textit{t} is (not the timestamp of the message itself, but) the timestamp of the last status change, and thus remains the same until the next change. 
An approximate timestamp value would be sufficient for messages to be accepted at their destination given that
{\it timeAllowedtoLive} is typically greater than 100 ms, much larger than the targeted latency of 1-2 ms.
%
%
%
The data conveyed in {\it allData} field consists of several binary values or a few multi-byte values to convey current, voltage, frequency readings depending on the type of dataset in the GOOSE message. Based on our observation from the SCL files of Electric Power and Intelligent Control (EPIC) testbed in the Singapore University of Technology and Design, there are three types of messages: 
Control, Protection, and Measurement. Control data includes two boolean values indicating circuit breaker status and a \textit{quality} value (generally ``0000'') associated with each. Similarly, the Protection data contains a boolean field indicating a fault occurrence, the same \textit{quality} value. The prediction of these binary values is viable given the limited space for possible outcomes. 

Measurement data involves 10-12 measurements, each consisting of several bytes representing voltage, current, or frequency.
Such measurements fluctuate within a certain range (e.g., around 49.9-50.1 Hz) and do not change markedly over time.
Using a prediction method (e.g., \cite{DASILVA2020106793}, \cite{da2017new}) can narrow down the space further. Thus, the set of possible measurements can be reduced to a much smaller set of Measurement packets to be prioritized. While predicting a large number of measurements is often non-trivial, Measurement packets are not as time-critical as Protection and Control packets. Thus, our scheme can still be opportunistically applied to Measurement packets with lower priority, as shown in Section \ref{sec:tree}.

\section{Baseline Caching-based Approaches} \label{sec:multicast}

\begin{figure}[ht]
\centering
\includegraphics[width=\linewidth,trim={0cm 0.9cm 0cm 0cm},clip]{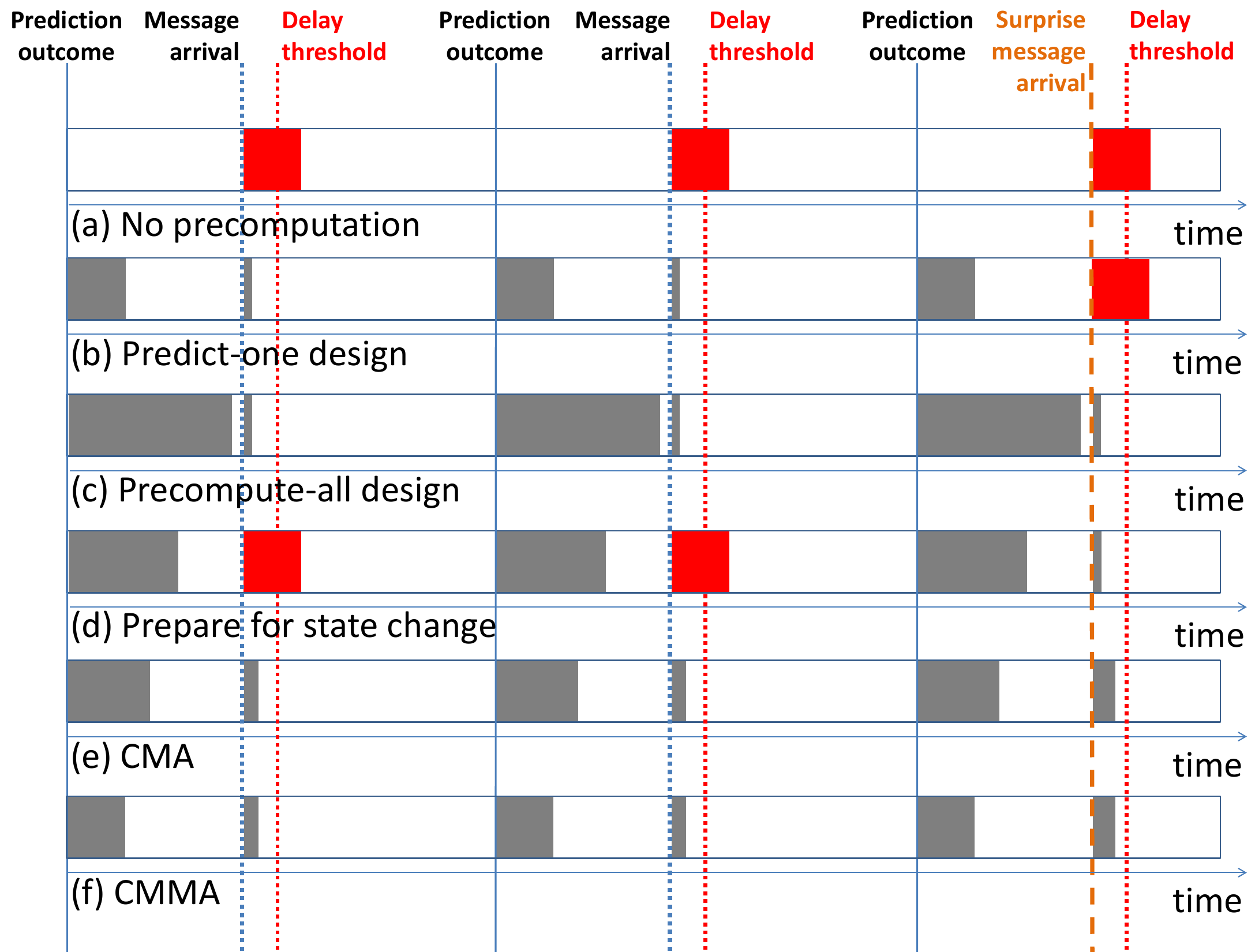}
\caption{Processing loads and packet delays under different designs. The scaling does not represent true values.}
\label{fig:multicast}
\end{figure}

To meet the stringent latency constraints (illustrated as `Delay threshold' in Fig.~\ref{fig:multicast}), a straightforward solution is to use lightweight cryptographic MAC without any need for message prediction. Such methods based on symmetric cryptography enable very fast message authentication in unicast settings, however, in the case of multicast or broadcast, the source needs to calculate cryptographic evidence separately for each recipient (using pairwise shared secret keys), which increases the computation time at the publisher linearly with the number of recipients. 
Such no precomputation setting is illustrated in Fig.~\ref{fig:multicast}(a), and its performance is given in Table \ref{tbl:multicast-analysis}. 

\begin{table*}[t!]
\caption{Complexity analysis. N: number of subscribers, k: number of unpredictable binary fields, k$_u$: number of unpredictable binary fields in an urgent message ($k_u<k$), R$_1$: message arrival rate, R$_2$: $1/${\it timeAllowedtoLive}, $D$: depth of the true message.}

\label{tbl:multicast-analysis}
\centering
\begin{tabular}{p{2.4cm}|p{2.2cm}|p{3cm}|p{2.2cm}|p{2.4cm}}
Design & \# of secure hashing in post-message & \# of secure hashing at the publisher & \# of secure hashing at the subscriber& communication overhead\\ \hline
No precomput.            & 2N         & 2N $\times$ R$_1$                         & 2R$_1$    & N $\times$ R$_1$ \\
Predict-one              & $\sim$0    & 2N $\times$ R$_2$                         & 2R$_1$    & N $\times$ R$_1$ \\
Precompute-all           & 0          & 2N $\times$ 2$^k$ $\times$ R$_2$          & 2R$_1$    & N $\times$ R$_1$ \\
State change             & $\sim$2N   & 2N $\times$ (R$_1+$2$^{k_u}\times$R$_2$)  & 2R$_1$    & N $\times$ R$_1$ \\

\hline

\brandNameShort\ w/ MHT  & 0       & (2N$+2^{k+1}$ $-1$) $\times$ R$_2$   &  ($k+$3) $\times$ R$_1$  & N $\times$ $(k+2)$ $\times$ R$_1$ \\
\brandNameShort\ w/ HHT  & 0       & (2N$+2^{k+1}$ $-1$) $\times$ R$_2$   &  ($D+$3) $\times$ R$_1$  & N $\times$ $(D+2)$ $\times$ R$_1$ \\ \hline
\brandName\ w/ MHT       & 0       & (3$+2^{k+1}$) $\times$ R$_2$         &  ($k+$5) $\times$ R$_1$  & ($k+$3) $\times$ R$_1$ \\
\brandName\ w/ HHT       & 0       & (3$+2^{k+1}$) $\times$ R$_2$         &  ($D+$5) $\times$ R$_1$  & ($D+$3) $\times$ R$_1$ \\ \hline

\end{tabular}
\vspace{-7.5pt}
\end{table*}



Based on the IEC 61850 GOOSE specification, messages are sent at arbitrary times whenever a change occurs or at regular intervals in the case of no change. The source can cache the cryptographic evidence within each {\it timeAllowedtoLive} period. If the source caches cryptographic evidence in anticipation of a state change, but no state change occurs within that {\it timeAllowedtoLive} period, the cache is not used and discarded. This results in a higher rate of cache build-up than the actual message arrival rate. Thus, the rate required to build cache for a message (R$_2$ = $1/timeAllowedtoLive$) would be greater than the actual message arrival rate (R$_1$). 

{\bf \noindent Predict-one design}: This design significantly reduces the average processing delay by caching the cryptographic evidence for only a single prospective message (preferably, the message with the highest probability to be sent), before the actual message is given. The second row of Table \ref{tbl:multicast-analysis} shows the required number of secure hash operations post-message. Fig.~\ref{fig:multicast}(b) illustrates the processor loads and packet delays for a case where the first two predictions hold true; hence the delays are minimal. However, the third prediction is wrong, and the evidence for the ``surprise'' message needs to be generated on-the-fly, and this will still incur a delay as large as the no precomputation setting. Predict-one design would only be suitable for systems that require low average communication delay, but can tolerate higher delay occasionally for surprise messages.

{\bf \noindent Precompute-all design}: If the source could cache the cryptographic evidence for all the possible prospective messages, it would avoid cryptographic operations in the post-message phase. As shown in Fig.~\ref{fig:multicast}(c), this design is feasible if precomputing a MAC for each possible message and destination is within the computing capability of the source hardware. This design ensures a delay upper bound at the cost of increased computation load. However, the computation load grows exponentially with the number of unpredictable binary fields in a message.



\noindent \textbf{Prepare for state change}: As discussed in Section \ref{sec:GOOSEdisc}, the periodic reporting messages of GOOSE can typically tolerate some delay, while the timely delivery of urgent control messages (triggered by an unexpected event such as a circuit breaker failure) is critical. In this design, the source always caches the cryptographic evidence only for such urgent messages. Although the load for precomputing for the urgent messages still grows exponentially with the number of binary fields in an urgent message ($\sim$N $\times$ 2$^{\textnormal{k}_u}$ $\times$ R$_2$), this is still lower than the precompute-all design, since such urgent messages report a smaller number of unpredictable binary fields, i.e., k$_u<$ k. 
The first two messages in Fig.~\ref{fig:multicast}(d) are periodic messages containing no state change. Therefore, the source has to generate the cryptographic evidence for the actual message after its arrival, incurring a delay equivalent to that of the no precomputation setting. When an urgent message arrives as in the third one, the delay is much smaller because the cryptographic evidence has already been cached.

Although the baseline approaches can be useful in certain settings, they rely on heavy precomputation to reduce message authentication delay. The precomputation and communication loads increase dramatically with the number of unpredictable binary fields and the number of destinations. 
Next, we discuss the use of authenticated trees and their integration with delayed key disclosure, which results in smaller loads and delays as illustrated in Figure~\ref{fig:multicast}(e)-(f).

\section{Caching-based (Multicast) Message Authentication}
\label{sec:tree}

Using authenticated data structures, \brandNameCommon\ reduces the precomputation load and communication overhead. Instead of computing cryptographic evidence separately for each message, the source constructs a binary tree on the set of prioritized messages, and uses the tree's root as an aggregate prioritization outcome. This root is shared with the destination(s) in the pre-message phase. Thus, instead of sharing each possible or prioritized future message and their corresponding evidence separately, the source caches the binary tree, then shares its root (and a proof to authenticate it), which serves as a public meta-data to authenticate the true message, provided that the true message is among the set of prioritized messages. We present the use of MACs, as well as an adaptation of TESLA protocol~\cite{perrig2002tesla} to authenticate the root.

\begin{figure}[t!]
\centering
\vspace{-0pt}
\includegraphics[width=\linewidth,trim={0.03cm 0.03cm 1.3cm 0.03cm},clip]{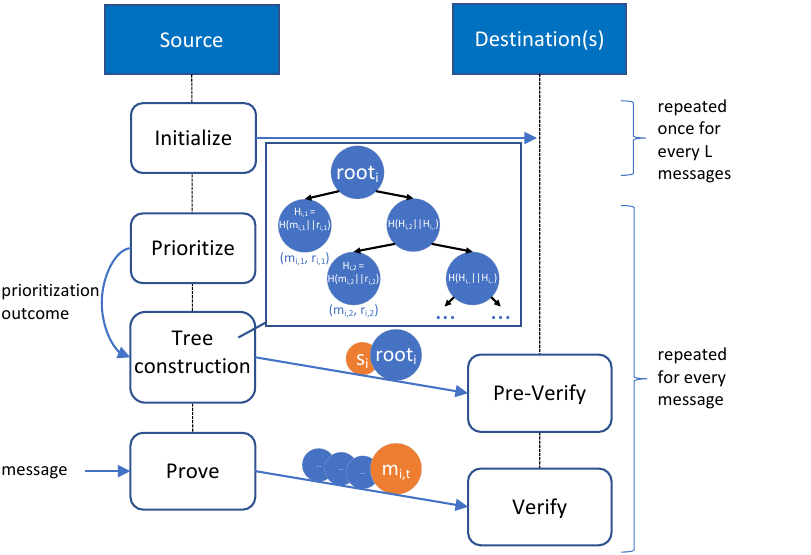}
\caption{\brandNameCommon\ model. $m_{i,1}$ is the most delay stringent (or likely) message, followed by $m_{i,2}$.}

\label{fig:protocolAndHHT}
\vspace{-15pt}
\end{figure}

Analogous to the construction of minimum redundancy codes \cite{huffman1952method}, Huffman Hash Tree (HHT) is constructed on the possible messages, in a way that the expected depth of the leaf corresponding to the true message is minimized. 
To achieve this, the depth of the leaf corresponding to a certain message is set to the equivalent of the coding length of that message in Huffman Coding. As illustrated in Figure~\ref{fig:protocolAndHHT}, this setup situates the prioritized messages (e.g., $m_{i,1}$) closer to the root. Since Merkle Hash Tree (MHT) is a special case of HHT where the frequencies or delay tolerance profiles of different messages are equal, we provide our example over an HHT. In an HHT implementation, the most likely messages, such as the expected measurements, or alternatively the most delay stringent messages (e.g., certain alerts) can be placed closer to the root. 

\subsection{\brandNameShort\ for Unicast ICS Communication} 
\label{sec:PreMA}
In the following, we describe the protocol using MACs to authenticate the root.

{\bf \noindent Initialize}($\lambda$)$\longrightarrow$\{$\mathcal{K}$\} :
In this procedure, given the security parameter $\lambda$, the source establishes pairwise symmetric keys $K=\{sk_1,\ldots,sk_N\}$ with each destination. This procedure is performed only during initialization, and repeated for every $L$ messages.

{\bf \noindent Prioritize}(preferences, system data)$\longrightarrow$\{$M_i$, $P_i$\}:
The input to the Prioritize procedure includes operator's preference (e.g., in terms of message type or target devices to prioritize) and historical system data.
The output consists of the set of possible messages $M_i=\{m_{i,1},\ldots,m_{i,2^k}\}$, and (if available) the normalized weight of each message $P_i=\{p_{i,1},\ldots,p_{i,2^k}\}$ (based on probabilities or tolerable delays), such that $\sum_{j=1}^{2^k} p_{i,j}$=1, for time interval $i\in\{1,\ldots L\}$. The output \{$M_i$, $P_i$\} is collectively referred to as the prioritization outcome. Whenever \textit{timeAllowedtoLive} is about to expire, or a new input is available, the procedure is repeated.

{\bf \noindent Tree construction}($M_i$, $P_i$, $K$, $ts_i$)$\longrightarrow$\{tree$_i$, root$_i$, $S_i\}$:
The inputs to this procedure are $M_i$, $P_i$ obtained from the output of the Prioritize procedure, the symmetric keys $K$ shared with each destination, and the timestamp $ts_i$ for freshness. The outputs are the tree tree$_i$, its root value root$_i$, and the set of HMACs $S_i = \{s_{i,1},s_{i,2},\ldots,s_{i,N} \}$ calculated from root$_i$, using the corresponding shared key of each destination $n\in \{1,\ldots,N\}$. The constructed tree binds the prioritized messages to a root value root$_i$. To prevent an adversary from predicting the messages and hence calculating the same root value, each message is concatenated with a nonce in $\{r_{i,1},\ldots,r_{i,{2^k}}\}$ before calculating the leaf values. Then, the root is obtained by pairwise hashing of sibling nodes starting from the deepest nodes. Finally, the root is timestamped and shared with the destination(s) along with its HMAC, calculated separately for each destination using the pairwise symmetric keys established in the Initialize procedure. In other words, $s_{i,n} =$ $HMAC(sk_n,ts_i,\textnormal{root}_i)$ is shared with destination $n$, $\forall n\in\{1,\ldots,\textnormal{N}\}$, where $sk_n$ is the key shared between the source and destination $n$, and $ts_i$ is the timestamp for the corresponding interval. The tree is only known to the source at this point, thus serves as the private meta-data.

{\bf \noindent Prove}($m_{i,t}$, tree$_i$)$\longrightarrow$\{proof$_i$\}:
The inputs are the true message $m_{i,t}$, and the $tree_i$ constructed in the previous procedure, and the output proof$_i$ is a collection of values from the tree between $m_{i,t}$ and the root. After the true message (say $m_{i,t}$) is known to the source, it sends $m_{i,t}$ and the other corresponding values of the HHT as proof$_i$ to allow the destination(s) to calculate the root value. For example, if $m_{i,t} = m_{i,2}$ in Fig.~\ref{fig:protocolAndHHT}, the proof contains $\{r_{i,2},H_{i,1}, H_{i,3}, H(H_{i,4} \| H_{i,.}) \}$. No hash operations are performed in this procedure.

Next, we introduce the steps at the destination device(s). 

{\bf \noindent Pre-Verify}($sk_n$, $s_{i,n}$, root$_i$)$\longrightarrow$\{accept, reject\}:
For destination $n$, the inputs are the shared key $sk_n$, the root root$_i$ and its HMAC $s_{i,n}$ sent by the source following its Tree Construction procedure.
When the destination $n$ receives the root value root$_i$ and its claimed HMAC $s_{i,n}$, it verifies root$_i$ using the shared key $sk_n$ established in the  Initialize procedure. If accepted, root$_i$ is stored for a \textit{timeAllowedtoLive} period. Note that this pre-verification is done before the actual message is given to the destination device.

{\bf \noindent Verify}($m'_{i,t}$, proof$_i$, root$_i$)$\longrightarrow$\{accept, reject\}:
The inputs are the message $m'_{i,t}$ to be verified, its proof proof$_i$, sent by the source following its $Prove$ procedure and the stored root value root$_i$ corresponding to the i$^{th}$ interval. Once the message $m'_{i,t}$ and its proof (corresponding values of the HHT) are received, the destination calculates the root by traversing the tree to retrieve hashes of siblings of nodes on a path between the leaf holding $m'_{i,t}$ and a root of the hash tree. Finally, the destination compares the calculated root with the previously stored roots which had been received within the last \textit{timeAllowedtoLive} duration. If a match is found, $m'_{i,t}$ is authenticated.

\subsection{\brandName\ for Broadcast/Multicast ICS Communication} \label{sec:premmaTESLA}

Using binary trees to combine the prioritization outcome into a single root value reduces the communication overhead and the computation overhead of the source. However, the factor of $N$ still appears in the corresponding complexity expressions in Table \ref{tbl:multicast-analysis}, because the cryptographic evidence to prove the root integrity is generated separately for each destination. In a multicast setting, to avoid computing MACs separately for each destination, we need to introduce a source of asymmetry between the source and destinations. The asymmetry ensures that the destinations can only verify the prioritization outcomes, but not generate valid evidence for them. We use an adaptation of TESLA protocol \cite{perrig2002tesla} to introduce time asymmetry while relying on the reasonable assumption that the destinations are loosely time synchronized with the source in a smart grid.\\

\vspace{-7.5pt}
\noindent\textbf{The plain TESLA protocol:}
We first discuss how the plain TESLA protocol operates. The source generates a hash chain by iteratively applying a one-way function $H$ --- constructed using a pseudorandom function family --- starting from a random number $C_L$. That is, $C_{i-1}=H(C_{i})$, $\forall i\in{1,\ldots,L}$ hence producing the sequence, $C_0, C_1, \ldots, C_L$ in the reverse order of generation. Since $H$ is one-way, no user other than the source knows $C_{i}$ given $C_{i-1}$. However, any user possessing $C_{0}$ can readily verify if a given $value$ belongs to the hash chain (and hence generated by the source) by checking if $H^i(value)=C_{0}$ for some $i$. 

After generating the hash chain, the source distributes $C_0$ to every destination securely, e.g., using digital signatures, or using the commitment of the previous hash chain ($C_L$) if any. To authenticate each message, the source computes and attaches the MAC using the key chain in the reverse of generation: $C_1, C_2, \ldots, C_L$. I.e., for the $j$'th message, $C_j$ is used.\footnote{In TESLA protocol another one way function $H'$ --- derived in the same way as $H$ --- is applied on $C_i$'s to derive the actual keys used in MAC computation. This is because using the same key both to derive the hash chain and to compute MACs may lead to cryptographic weaknesses.} Also along with the $j$'th message, the source reveals the key $C_i$, which was used to compute the MAC of the earlier $i$'th message, so that the destination(s) can verify the authenticity of message $i$ by checking $H^i(C_i)=C_0$, or simply $H(C_i)=C_{i-1}$ if the destination possesses $C_{i-1}$. $d=j-i$ number of messaging intervals is the disclosure delay of TESLA, and ensures source-destination(s) asymmetry without having to resort to expensive public key cryptography.

In TESLA, one MAC per message is sufficient to provide broadcast/multicast authentication, provided that the destinations have loose time synchronization with the source. However, the major drawback of TESLA (or delayed key disclosure schemes in general) is, the delay in verification of each message, introduced by the disclosure delay. Clearly, disclosure delay in the order of a few messaging intervals (i.e., larger than the sum of maximum network delay and synchronization error) is not tolerable in the time-critical setting we are targeting.\\ 

\begin{figure}[h]
\centering
\vspace{-7.5pt}
\includegraphics[width=\linewidth,trim={0.05cm 0.05cm 0.02cm 0.1cm},clip]{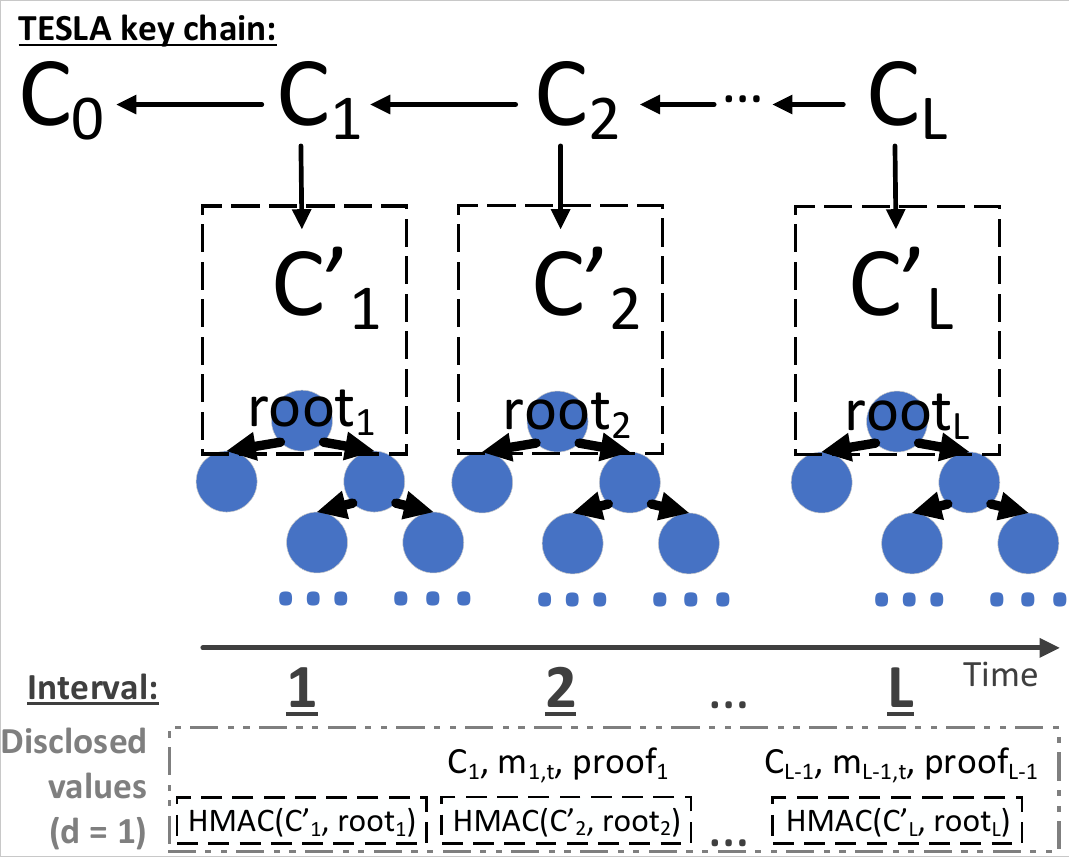}
\caption{The disclosure schedule of \brandName .}
 \label{fig:LMF}
\end{figure}

\vspace{-7.5pt}
\noindent\textbf{\brandName :}
\brandName\ relies on authenticating the prioritization outcomes with TESLA, rather than the true message itself, as shown in Fig.~\ref{fig:LMF}. Therefore, it does not suffer from the disclosure delay of TESLA, despite using it to introduce source-destination asymmetry. Next, we describe the procedures.

{\bf \noindent Initialize}($C_{L}'$)$\longrightarrow$\{$C_0,\ldots,C_{L}$, proof for $C_0$\}:
This procedure outputs a key chain of $L+1$ values, and the proof for the first value by using the commitment of the previous hash chain, $C_{L}'$. The source selects a random number $C_L$, and commits it to generate a hash-chain of length $L+1$, by repeatedly applying a one-way function $H$, such that $C_{i-1}=H(C_{i})$, $\forall i\in{1,\ldots,L}$. We call $\{C_i\}$'s as the TESLA keys and they are used in the reverse order of generation, i.e., from $C_1$ to $C_L$, to authenticate the root values for the messages in the next $L$ time intervals. Finally, the source generates a proof for the first TESLA key $C_0$ and the key disclosure schedule using the commitment of the previous hash chain ($C_{L}'$) if any, otherwise digitally signs $C_0$ and the key disclosure schedule before sending them to the destination(s).

As stated above, in TESLA another hash function $H'$ is applied on $C_i$'s to derive the actual keys used in MAC computation. Thus, we follow the same rule and use another set of keys derived from the TESLA keys, $C_i' = H'(C_i)$, $\forall i\in{0,\ldots,L}$, to compute the MAC for messages in the next $L$ time intervals.

{\bf \noindent Prioritize}(preferences, system data)$\longrightarrow$\{$M_i$, $P_i$\}: This procedure is the same as the Prioritize procedure of \brandNameShort\ above.


{\bf \noindent Tree Construction}($M_i$, $P_i$, $C_{i}$, $ts_i$)$\longrightarrow$\{tree$_i$, root$_i$, $s_i$\}: The inputs to this procedure are $M_i$, $P_i$ obtained from the output of the Prioritize procedure, the TESLA key $C_{i}$ for time interval $i$, and the timestamp $ts_i$. The outputs are the tree tree$_i$, its root value root$_i$, and the MAC $s_i$ for $root_i$, computed using $C_{i}$. In this procedure, the source constructs an HHT on the set of prioritized messages. Since an adversary can also predict such messages, each leaf node should be nonced with a random value in $\{r_{i,1},\ldots,r_{i,{2^k}}\}$. 
The source then computes a MAC for the tree's root, root$_i$, using the hash of the corresponding TESLA key $C_{i}$, such that $s_i = HMAC(H'(C_i),ts_i,\textnormal{root}_i)$. $s_i$ is shared with the destination(s) and serves as the public meta-data. The tree is only known to the source, thus serves as the private meta-data. Note that only a single public meta-data value $s_i$ is computed for all destinations.

{\bf \noindent Prove}($m_{i,t}$, tree$_i$)$\longrightarrow$\{proof\}:
The inputs are the true message $m_{i,t}\in M_i$, and the $tree_i$ constructed in the previous procedure, and the output proof$_i$ is a collection of values from the tree. After the true message $m_{i,t}$ is known to the source (say $d$ time intervals later), it sends $m_{i,t}$ and the corresponding values of tree$_i$ that will allow the destination(s) to calculate root$_i$. The source also discloses the TESLA key $C_i$ together with the message to allow the destination(s) to verify root$_i$. Note that the TESLA key $C_i$ used to verify the root$_i$ is disclosed in time interval $d+i$ to create time asymmetry between the source and destination(s), but this disclosure delay is not reflected in the authentication of $m_{i,t}$, as long as the prioritization outcome precedes the true message by more than the minimum disclosure delay allowable by the system.\footnote{The disclosure delay depends on the level of time synchronization, and the network delay between the source and destination(s). Typically, the disclosure delay should be the sum of the network delay and synchronization error \cite{perrig2000efficient}.}

{\bf \noindent Pre-Verify}: This step is null in \brandName.

{\bf \noindent Verify}($C_{i},C_{i-1}, s_i, m_{i,t}$, proof$_i$)$\longrightarrow$\{accept, reject\}: The inputs are TESLA keys $C_{i},C_{i-1}$, the MAC $s_i$ of $root_i$, the true message $m_{i,t}$ and the proof proof$_i$ for the true message. In this procedure, first, the destination verifies the TESLA key $C_{i}$ using a previously disclosed key, e.g., $H(C_{i})=C_{i-1}$. If $C_{i-1}$ was not received due to packet loss etc., the destination can still verify $C_{i}$ by repeatedly applying $H$ on $C_{i}$ to obtain the last received TESLA key, i.e., $H^j(C_{i})=C_{i-j}$. If verified, $C_{i}$ is used to verify the true message, and also stored to verify the future values of the hash chain to be received. Then, using the proof$_i$, the destination traverses the tree to retrieve hashes of siblings of nodes on a path starting from the leaf holding $m_{i,t}$, hence calculates the root value. Finally it verifies if $s_i = HMAC(H'(C_i),ts_i,\textnormal{root}_i)$. If verified, the true message is authenticated.


\subsection{Performance and Security Analysis}

We discuss the complexity of \brandNameShort\ and \brandName\ over Table~\ref{tbl:multicast-analysis}. In both schemes the source does not perform any computations other than memory reads and packet assembly in the post-message phase. It simply collects the corresponding values from the tree and piggybacks them to the message. For the pre-message phase, given $k$ possible binary fields and hence $2^k$ possible messages in a prioritization outcome, the binary tree can be constructed with $2^{k+1}-1$ hash operations (i.e., $2^k$ to generate the leaves, plus $2^k-1$ to construct the rest of tree). Given prioritization outcomes at a rate of R$_2$, the required computing rate at the source to generate the binary tree would be $({2^{k+1}-1})$R$_2$ number of secure hash operations per unit time. The tree generation complexity is common for both \brandNameShort\ and \brandName . Added to this is the generation of MACs, which requires $2\textnormal{N}$R$_2$ secure hashing operations (two per destination, per tree) for \brandNameShort\ and $4$R$_2$ secure hashing operations (two for the MAC of root, one for deriving the TESLA key and one for the key of MAC from the TESLA key) for \brandName\ per unit time. So the total computing load at the source  would be $({2\textnormal{N}+2^{k+1}-1})$R$_2$ for \brandNameShort\ and ($3+2^{k+1}$) $\times$ R$_2$ for \brandName . 

In \brandNameShort\ the destination verifies the root before (Pre-Verify), and the proof after (Verify) the message is received. The root verification in the Pre-Verify procedure uses MACs, and demands a computing rate of $2$R$_1$ secure hashing both for MHT and HHT variants. For a MHT, proof verification in the Verify procedure costs $(k+1)$R$_1$ secure hashing operations. The verification complexity of the HHT proof depends on where the received message is located on the tree. The required computing load is $2$R$_1$ for the message at depth $1$, and $(D+1)$R$_1$ in general, where $D$ is the depth of the actual message in the HHT, and $D\leq2^k-1$. So, the total computing load at the destination is between $4$R$_1$ and $(2+2^k)$R$_1$ for HHT and $(k+3)$R$_1$ for MHT variants. In \brandName\, in addition to the computations above, the corresponding TESLA key is verified using the previously disclosed TESLA key, and the key of the MAC is generated from it, each costing a secure hashing. Therefore the computing load is $2$R$_1$ more than the \brandNameShort\ variant.

The proof size is $(k+1)$ hash values for MHT, and $(D+1)$ for HHT, which yields 2 for the most likely (or delay stringent) message in \brandNameShort . Adding the MAC to the proof contributes one more hash value to the communication overhead. In \brandNameShort\ the proof is communicated separately to $N$ destinations, hence bringing the total communication overhead to $N(k+2)$ and $N(D+2)$ for MHT and HHT variants. In \brandName\ the same proof and MAC, as well as the corresponding TESLA key, are shared with all destinations, totaling $N(k+3)$ and $N(D+3)$ hash values for its MHT and HHT variants.

In \brandNameCommon\ a nonce is released when the message corresponding to that nonce is sent. Therefore, a nonce should be refreshed once that happens. If a probabilistic polynomial time adversary can find any set of values that give the same root value without knowing the nonces, we can either break the preimage resistance or the collision resistance of the hash function employed. For 128-bit security, one should use 256-bit nonces and a secure hash function, such as SHA-256. 

\subsection{Security Proof Sketch}
In line with the threat model in Section \ref{sec:GOOSEdisc}, we consider an adversary that forges a signature for any message of his choice, where the message does not necessarily have any particular format or meaning, but has never been signed by the legitimate source. Hence, we consider the existential forgery game under adaptive chosen-message attack.

There are two steps to consider in the proof of security; one for the secure transmission of the root (binding the aggregate prioritization outcome) and one for the authentication of the true messages. The first is as secure as the employed MAC (for \brandNameShort) or TESLA (for \brandName). For the second, we show that our scheme is secure if the hash function employed for the authenticated data structure is collision and preimage resistant. 

If a probabilistic polynomial time (PPT) adversary $\mathcal{A}$ wins security game of our scheme with non-negligible probability, we use it to construct other PPT algorithms $\mathcal{B}$ or $\mathcal{B}'$ who break with non-negligible probability, the collision resistance or the preimage resistance of the hash function, respectively. 
$\mathcal{B}$ acts as the adversary in the security game with the hash function challenger $\mathcal{HC}$. In parallel, $\mathcal{B}$ plays the role of the challenger in our game with $\mathcal{A}$. Consider the following existential forgery game under adaptive chosen-message attack:

\noindent \textbf{Setup:} $\mathcal{HC}$ picks a secure hash function (\textit{hash}) from a hash function family --- we use SHA-256 in our implementation --- and passes the parameters to $\mathcal{B}$. $\mathcal{B}$ generates a set of nonce values of size equivalent to the maximum possible size of the prioritized message set. 

\noindent \textbf{Query:} $\mathcal{A}$ generates a set of values (messages and relative probabilities) and passes it to $\mathcal{B}$ to be used as the prioritization outcome. $\mathcal{B}$ builds a tree\footnote{Each tree is of size equivalent to the maximum number of prioritized messages, which is bounded by a constant.} on the prioritization outcome. 
In particular, $\mathcal{B}$ calculates the hash values using \textit{hash} for every node of the tree and passes the root hash value ($root_i$ for the $i$th query) to $\mathcal{A}$. $\mathcal{B}$'s load is constant in this step, given that each prioritization outcome is of constant size $c$ and creating the tree is linear in $c$ (specifically, $2c$). $\mathcal{A}$ then picks a message $m_i$ among the set of generated messages for the $i$th query, and passes it to $\mathcal{B}$. $\mathcal{B}$ performs the Prove procedure using $m_i$ as the true message, hence collects the nonce value from the corresponding leaf and the neighboring values on the path from the leaf to the root (collectively called $proof_i$ for query $i$), then passes them to $\mathcal{A}$. $\mathcal{A}$ can repeat this query for polynomially many ($p$) times. After query $i$, $\mathcal{B}$ refreshes the consumed nonce, which is revealed to generate $proof_i$.

\noindent \textbf{Challenge:} $\mathcal{A}$ prepares and sends another message $m'$ where $m'$ $\neq$ $m_i$, $\forall i \in \{1,\ldots, p\}$ and a set of values (proof$'$) that are required to calculate the hash value of the root $root_i$, in any query $i \in \{1,\ldots, p\}$ that $\mathcal{A}$ picks.

If $\mathcal{B}$ verifies the proof$'$ and $m'$ with $root_i$ for any $i \in \{1,\ldots, p\}$, using the Verify procedure, then $\mathcal{A}$ wins. For this, $\mathcal{A}$ needs to find a nonce value that has not been revealed to it or needs to come up with a new set of values, which, when chained together with the given hash function \textit{hash}, yield any $root_i$. For the former, the probability that $\mathcal{A}$ finds a nonce that has not been revealed to it is negligible, since otherwise, we can define a $\mathcal{B}'$ that breaks the preimage resistance of \textit{hash}. For the latter, $\mathcal{B}$ uses the new set of values to break the security of the given hash function (by finding a collision in \textit{hash}) using the collision on the chain from the leaf node corresponding to $m'$ to the root.

\section{Implementation and Evaluation}\label{sec:evaluation}

We used BeagleBoard-X15 (BBX15 for short) with a single active core for measuring the computation times in the schemes discussed in this paper. The reason for choosing BBX15 is its relatively low-cost, and its port selection that accommodates a Bump-in-the-wire (BITW) deployment, a desirable feature for legacy compliance. 
Detailed specifications of BBX15 can be found in \cite{beagleboard}. The HHT generated in the Tree construction steps of \brandNameCommon\ is stored in the RAM. All data points presented are the average of 500 runs. The security parameters for each implemented scheme are chosen to meet 128-bit security, in particular, SHA-256 as the digest function, 256-bit nonces, and curve P-256 in ECDSA.

We have measured the average time taken for subtasks of the baseline designs, \brandNameShort\ and \brandName, as well as the widely used ECDSA. For ECDSA plots, we assumed the same authenticated tree based approach to bind prioritization outcomes into a single root value, but used ECDSA to authenticate the root. In other words, the source signs the tree's root with ECDSA, instead of generating a MAC for the root in the Tree construction procedure as in \brandNameCommon. Consequently, the Pre-Verify step of ``tree with ECDSA'' involves the verification of ECDSA signature. 
Since the post-message signing consists only of memory reads and packet assembly (see the Prove procedures of \brandNameShort\ and \brandName\ in Sections \ref{sec:PreMA} and \ref{sec:premmaTESLA}), it is performed virtually instantly, and hence not evaluated. 

\subsection{Precomputation Time}
\begin{figure}[h]
\centering
\includegraphics[width=\linewidth]{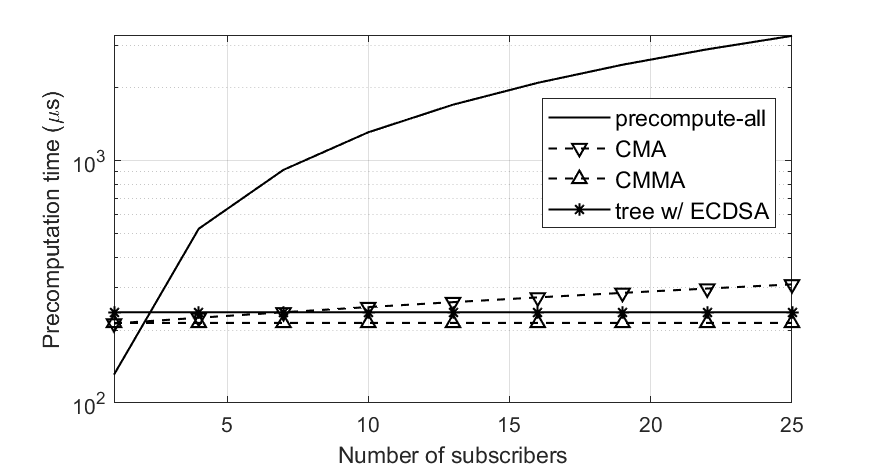}
\caption{Precomputation time over the number of subscribers.}
\label{fig:precomp1}
\end{figure}

The precomputation time does not contribute to the delay overhead as long as it is shorter than a messaging interval, yet it determines the maximum messaging throughput. Fig.~\ref{fig:precomp1} compares \brandNameShort\ and \brandName\ with ``tree with ECDSA'' and the baseline approach of precompute all, over the number of destinations, assuming 32 prioritized messages.  
As can be seen, the precomputation time increases with the number of destinations for precompute-all, and \brandNameShort\ schemes. This is because the source needs to precompute a separate proof for each destination, using the corresponding pairwise symmetric keys. Nevertheless, the amount of increase in \brandNameShort\ is much smaller thanks to the use of authenticated trees rather than generating a separate proof for each prioritization outcome. The precomputation times in \brandName\ and ECDSA do not depend on the number of destinations due to the time asymmetry for the former, and key asymmetry for the latter. \brandName\ outperforms ECDSA thanks to the use of symmetric keys. 

\begin{figure}[h]
\centering
\vspace{-7.5pt}
\includegraphics[width=\linewidth]{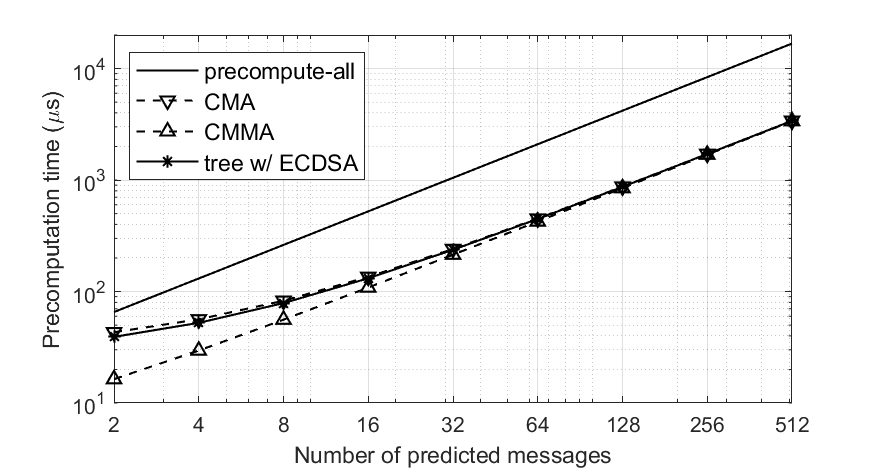}
\caption{Precomputation time over the number of messages.}
\label{fig:precomp2}
\vspace{-7.5pt}
\end{figure}

In Fig.~\ref{fig:precomp2}, precomputation time is plotted over the number of predicted/prioritized messages. All schemes suffer increased precomputation time due to multiple MAC computations in precompute-all and due to larger tree size in tree-based schemes (i.e., \brandNameShort, \brandName, and tree with ECDSA). At 32 messages, the precomputation times for \brandName\ and \brandNameShort\ are 210-250 $\mu$s. If the true message is always in the set of prioritized messages, \brandNameCommon , therefore potentially supports the throughput of 4000 messages per second in IEC 61850 SV. \brandName 's precomputation time is the shortest, outperforming tree'd ECDSA with approximately 23 $\mu$s margin regardless of the number of subscribers or messages in Figures \ref{fig:precomp1} and \ref{fig:precomp2}. Although the 23 $\mu$s difference is not very large, the main drawback of ECDSA is its verification time, as discussed next.

\subsection{Verification Time}
Since the Prove procedure is virtually instant in \brandNameShort\ and \brandName\, the post-message verification time is the largest contributor to the authentication delay.

\begin{figure}[h]
\centering
\vspace{-15pt}
\hspace*{-2pt}\includegraphics[width=\linewidth]{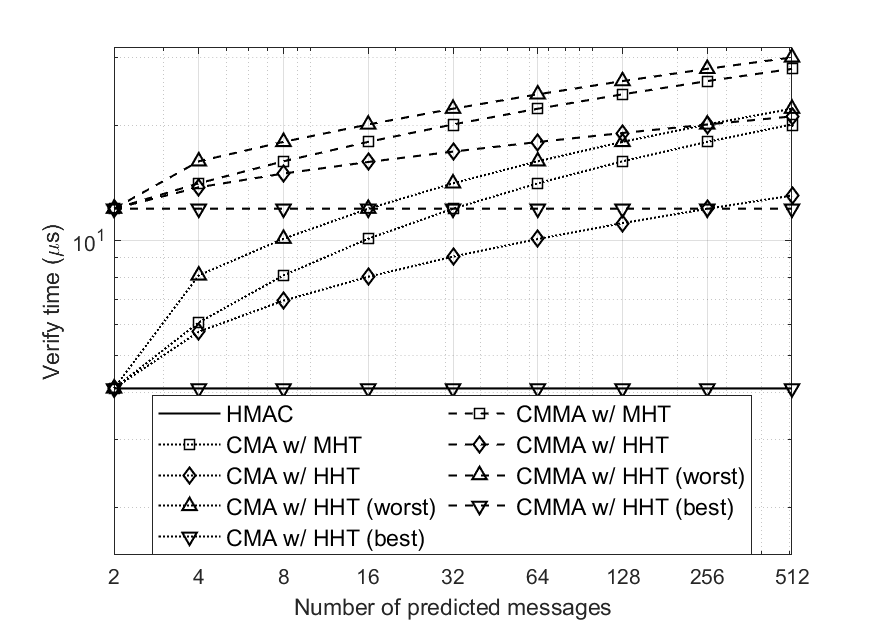}
\caption{Verification time over the number of messages. ECDSA is 2500 $\mu$s, therefore off the scale. Distribution 3 is assumed.}
\label{fig:verify1}
\vspace{-7.5pt}
\end{figure}

In Fig.~\ref{fig:verify1} we show the verification time of \brandName\ and \brandNameShort\ with the HMAC as a benchmark, since MACs (in straw man and precompute-all design) have the smallest verification time (4 $\mu$s). HHT based \brandName\ and \brandNameShort\ constructions have lower average verification times than those with MHT because the average depth of the actual message is minimized in HHT (assuming message likelihood based HHT). The best case for HHT based construction would be when the most likely message is the true message. 

In the plain TESLA protocol, the disclosure delay is added to the verification time, therefore, it would incur significantly larger verification delay than \brandNameCommon . As a rough comparison, even under ideal circumstances, e.g., perfect source-destination synchronization, zero network delay, the highest rate of 4000 messages/second in IEC 61850 SV~\cite{schaub2009iec}, and a minimal disclosure delay of one interval, TESLA would incur 250 $\mu$s compared to several $\mu$s of \brandNameCommon .

In Fig.~\ref{fig:verify2}, we illustrate how the probability distribution of future messages affects the verification time of \brandNameCommon\ with HHT. We only show \brandName\ on this figure for brevity (\brandNameShort\ would be 6-8 $\mu $s faster). Given that there are $2^k$ possible messages, the four probability distributions we consider are: 

\begin{figure}[h]
\centering
\includegraphics[width=\linewidth]{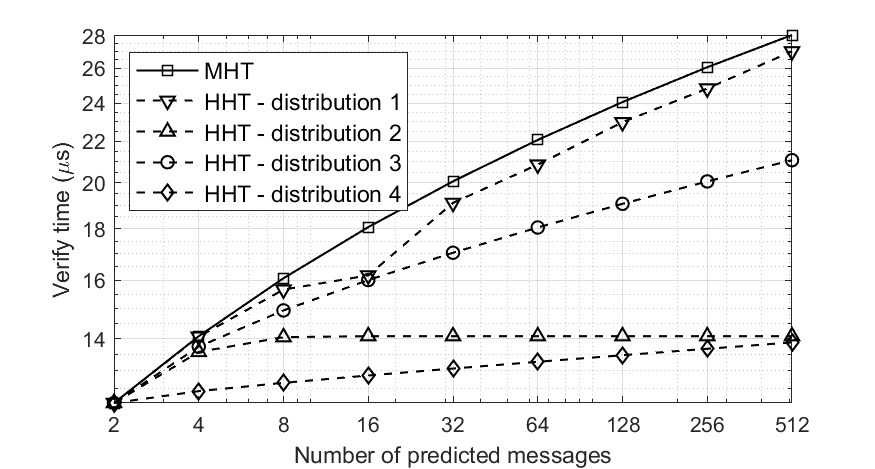}
\caption{Verification time under different probability distributions for prioritized messages.}
\vspace{-7.5pt}
\label{fig:verify2}
\end{figure}

\begin{enumerate}[wide, labelwidth=!, labelindent=0pt]
    \item [Distribution 1:] We draw samples of $2^k$ i.i.d exponential random variables and normalize their sum to $1$
    \item [Distribution 2:] {\small $\Pr(m_i) = 2^{-i}$, for $i\leq 2^k-1$, $\Pr(m_{2^k}) = 2^{-2^k+1}$}
    \item [Distribution 3:] {\small$\Pr(m_1) = 0.5$, $\Pr(m_i) = 0.5/(2^k-1)$ for $i \geq 2$}
    \item [Distribution 4:] {\small$\Pr(m_1) = 0.9$, $\Pr(m_i) = 0.1/(2^k-1)$ for $i \geq 2$}
\end{enumerate}


As can be seen from Fig.~\ref{fig:verify2}, \brandName\ with HHT performs significantly better than \brandName\ with MHT, when certain messages have a markedly higher probability than others (distributions 2 and 4).
In particular, distribution 4 yields the best results, because message 1, with a much lower depth in the HHT, is the message to be authenticated in 9 out of every 10 messaging intervals. This is also in line with IEC 61850 GOOSE/SV messaging, where the probability of sending a `no state change' message is much larger than any other.

\section{Related Work}\label{sec:related_work}

While digital signatures are widely used for multicast/broadcast authentication, they are not suitable for resource-constrained devices or delay stringent applications. Numerous schemes have been proposed, e.g., \cite{shamir2001improved,reyzin2002better,Aysu2016,ateniese2013low}, to offload some computations of digital signatures to a phase before the message is given. The first phases of El-Gamal~\cite{elgamal1985public}, DSS \cite{gallagher2013digital}, and precomputation enabled ECDSA~\cite{johnson2001elliptic} do not require the true message similar to our Tree Construction procedure. Online/offline signatures, either based on one-time signature schemes~\cite{even1989line} or based on chameleon commitments \cite{krawczyk2000chameleon,shamir2001improved,yang2020lis} can transform any digital signature scheme to a one with such offloading feature. The idea is to perform expensive cryptographic operations in the offline phase and generate a meta-data, so that in the online phase the source can sign a message only using inexpensive computations. Although the online signing in online/offline converted signature schemes can be fast, these schemes still fall short in meeting stringent latency requirements due to their offline phase or require a large volume of metadata to be stored by the source. Among these, Lis~\cite{yang2020lis} is specifically designed for cyberphysical systems. However, it is optimized for the publisher, and thus, the verification cost is still high. Several other schemes that amortize a signature over several packets, e.g., \cite{ParkAmortization,liu2012pkc,tartary2011authentication}, are also not satisfactory in avoiding large computation and communication overhead, not robust against packet losses, or lack immediate verification. The signature scheme in ~\cite{yavuz2014efficient} exploits the structure of ICS messages to enable precomputation for a verifier-efficient scheme such as RSA, but is limited only to certain commands in ICSs.

Another promising domain for lightweight message authentication is one-time signatures (OTS), such as those by Lamport \cite{lamport1979constructing} and Rabin \cite{rabin1979digitalized}. While these earlier OTS require large keys and signatures, Winternitz OTS (WOTS) Scheme \cite{merkle1989certified} reduced the signature size by trading off setup and verification time. In the same work, Merkle proposed the conversion of WOTS into a many-time signature, by constructing a Merkle Hash Tree (MHT) on multiple WOTS public keys to bind them into a single root value, which serves as the public key for multiple WOTS instances. Using MHT in OTS/WOTS enables fast signing of multiple messages with a single public key. More efficient variants of OTS such as BiBa~\cite{perrig2001biba} and HORS~\cite{reyzin2002better} have also been proposed. Nevertheless, the public key size in OTS and their BiBa/HORS variants is relatively large, and each public/private key pair can be used only once, rendering them impractical for bandwidth or storage constrained networks/devices.


\section{Conclusions}\label{sec:conclusion}
We proposed \brandNameShort\ and its multicast variant \brandName\ as  caching-based message authentication schemes that exploit the limited entropy of ICS messages to virtually eliminate the processing delay after the message is given. 
\brandNameCommon\ is suitable for ICS protocols such as IEC 61850 where the time-critical messages consist mostly of predetermined or predictable data. It relies on symmetric cryptography, therefore computationally efficient, and does not suffer from the disclosure delay of delayed key disclosure schemes, despite leveraging their time asymmetry. We have shown that \brandNameCommon\ is a lightweight alternative for digital signatures for low-entropy messages. Alternatively, \brandNameCommon\ can be opportunistically used in conjunction with digital signatures to lower the expected overhead of message authentication in ICSs.



\bibliographystyle{IEEEtran}
\bibliography{main}

\begin{thebibliography}{10}
\providecommand{\url}[1]{#1}
\csname url@samestyle\endcsname
\providecommand{\newblock}{\relax}
\providecommand{\bibinfo}[2]{#2}
\providecommand{\BIBentrySTDinterwordspacing}{\spaceskip=0pt\relax}
\providecommand{\BIBentryALTinterwordstretchfactor}{4}
\providecommand{\BIBentryALTinterwordspacing}{\spaceskip=\fontdimen2\font plus
\BIBentryALTinterwordstretchfactor\fontdimen3\font minus
  \fontdimen4\font\relax}
\providecommand{\BIBforeignlanguage}[2]{{%
\expandafter\ifx\csname l@#1\endcsname\relax
\typeout{** WARNING: IEEEtran.bst: No hyphenation pattern has been}%
\typeout{** loaded for the language `#1'. Using the pattern for}%
\typeout{** the default language instead.}%
\else
\language=\csname l@#1\endcsname
\fi
#2}}
\providecommand{\BIBdecl}{\relax}
\BIBdecl

\bibitem{ieee1646}
{IEEE Power and Energy Society}, ``{IEEE Standard Communication Delivery Time
  Performance Requirements for Electric Power Substation Automation},'' 2004.

\bibitem{fpro2019}
E.~Esiner, D.~Mashima, B.~Chen, Z.~Kalbarczyk, and D.~Nicol, ``F-pro: a fast
  and flexible provenance-aware message authentication scheme for smart grid,''
  in \emph{Proceedings of IEEE SmartGridComm 2019}.\hskip 1em plus 0.5em minus
  0.4em\relax IEEE, 2019.

\bibitem{IEC61850original}
``Communication networks and systems in substations,'' [Online]. Available:
  \url{https://webstore.iec.ch/}, (Date last accessed on Jul. 21, 2021).

\bibitem{IEC61850}
R.~E. Mackiewicz, ``Overview of {IEC} 61850 and benefits,'' in \emph{2006 IEEE
  Power Engineering Society General Meeting}.\hskip 1em plus 0.5em minus
  0.4em\relax IEEE, 2006, pp. 8--pp.

\bibitem{cleveland2005iec}
Cleveland, ``{IEC} tc57 security standards for the power system's information
  infrastructure - beyond simple encryption,'' in \emph{2005/2006 IEEE/PES
  Transmission and Distribution Conference and Exhibition}, May 2006, pp.
  1079--1087.

\bibitem{RSAsignature}
R.~L. Rivest, A.~Shamir, and L.~Adleman, ``A method for obtaining digital
  signatures and public-key cryptosystems,'' \emph{Commun. ACM}, vol.~21,
  no.~2, p. 120–126, Feb. 1978.

\bibitem{johnson2001elliptic}
D.~Johnson, A.~Menezes, and S.~Vanstone, ``The elliptic curve digital signature
  algorithm ({ECDSA}),'' \emph{International journal of information security},
  vol.~1, no.~1, pp. 36--63, 2001.

\bibitem{hauser1997reducing}
R.~Hauser, T.~Przygienda, and G.~Tsudik, ``Reducing the cost of security in
  link-state routing,'' in \emph{Proceedings of SNDSS'97: Symposium on Network
  and Distributed System Security}.\hskip 1em plus 0.5em minus 0.4em\relax
  IEEE, 1997, pp. 93--99.

\bibitem{cheung1997efficient}
S.~Cheung, ``An efficient message authentication scheme for link state
  routing,'' in \emph{Proceedings 13th Annual Computer Security Applications
  Conference}.\hskip 1em plus 0.5em minus 0.4em\relax IEEE, 1997, pp. 90--98.

\bibitem{perrig2002tesla}
A.~Perrig, R.~Canetti, J.~D. Tygar, and D.~Song, ``The {TESLA} broadcast
  authentication protocol,'' \emph{Rsa Cryptobytes}, vol.~5, no.~2, pp. 2--13,
  2002.

\bibitem{perrig2000efficient}
------, ``Efficient authentication and signing of multicast streams over lossy
  channels,'' in \emph{Proceeding 2000 IEEE Symposium on Security and Privacy.
  S\&P 2000}.\hskip 1em plus 0.5em minus 0.4em\relax IEEE, 2000, pp. 56--73.

\bibitem{liu2004multilevel}
D.~Liu and P.~Ning, ``Multilevel $\mu${TESLA}: Broadcast authentication for
  distributed sensor networks,'' \emph{ACM Transactions on Embedded Computing
  Systems (TECS)}, vol.~3, no.~4, pp. 800--836, 2004.

\bibitem{liu2005practical}
D.~Liu, P.~Ning, S.~Zhu, and S.~Jajodia, ``Practical broadcast authentication
  in sensor networks,'' in \emph{The Second Annual International Conference on
  Mobile and Ubiquitous Systems: Networking and Services}.\hskip 1em plus 0.5em
  minus 0.4em\relax IEEE, 2005, pp. 118--129.

\bibitem{TESLA_infocom}
Y.~Huang, W.~He, K.~Nahrstedt, and W.~C. Lee, ``Dos-resistant broadcast
  authentication protocol with low end-to-end delay,'' in \emph{IEEE INFOCOM
  Workshops 2008}, 2008, pp. 1--6.

\bibitem{li2011performance}
J.~Li, Q.~Huang, F.-k. Hu, and S.~Jing, ``Performance testing on goose and msv
  transmission in one network,'' \emph{Energy Procedia}, vol.~12, pp. 185--191,
  2011.

\bibitem{roomi2020}
M.~M. Roomi, P.~P. Biswas, D.~Mashima, Y.~Fan, and E.-C. Chang, ``False data
  injection cyber range of modernized substation system,'' in \emph{Proceedings
  of IEEE SmartGridComm}.\hskip 1em plus 0.5em minus 0.4em\relax IEEE, 2020.

\bibitem{case2016analysis}
R.~Lee, M.~Assante, and T.~Conway, ``Analysis of the cyber attack on the
  ukrainian power grid,'' 2016.

\bibitem{CrashOverride}
``Crashoverride malware,'' [Online]. Available:
  \url{https://www.us-cert.gov/ncas/alerts/TA17-163A}, 2017, (Date last
  accessed on Aug. 18, 2017).

\bibitem{GOOSEMessageStructure}
C.~Kriger, S.~Behardien, and J.~Retonda, ``A detailed analysis of the goose
  message structure in an {IEC} 61850 standard-based substation automation
  system,'' \emph{International Journal of Computers, Communications and
  Control ({IJCCC})}, vol.~8, pp. 708--721, 2013.

\bibitem{biswas2019synthesized}
P.~P. Biswas, H.~C. Tan, Q.~Zhu, Y.~Li, D.~Mashima, and B.~Chen, ``A
  synthesized dataset for cybersecurity study of {IEC} 61850 based
  substation,'' in \emph{Proceedings of IEEE SmartGridComm}.\hskip 1em plus
  0.5em minus 0.4em\relax IEEE, 2019.

\bibitem{DASILVA2020106793}
L.~{Da Silva} and D.~Coury, ``Network traffic prediction for detecting ddos
  attacks in {IEC} 61850 communication networks,'' \emph{Computers and
  Electrical Engineering}, vol.~87, p. 106793, 2020.

\bibitem{da2017new}
L.~E. Da~Silva and D.~V. Coury, ``A new methodology for real-time detection of
  attacks in {IEC} 61850-based systems,'' \emph{Electric Power Systems
  Research}, vol. 143, pp. 825--833, 2017.

\bibitem{huffman1952method}
D.~A. Huffman, ``A method for the construction of minimum-redundancy codes,''
  \emph{Proceedings of the IRE}, vol.~40, no.~9, pp. 1098--1101, 1952.

\bibitem{beagleboard}
``{BeagleBoard-X15},'' [Online]. Available: \url{https://beagleboard.org/x15}",
  2018, (Date last accessed on May 17, 2018).

\bibitem{schaub2009iec}
P.~Schaub and A.~Kenwrick, ``An {IEC} 61850 process bus solution for
  powerlink's ipass substation refurbishment project,'' \emph{PAC World
  Magazine}, vol.~9, no. 2009, pp. 38--44, 2009.

\bibitem{shamir2001improved}
A.~Shamir and Y.~Tauman, ``Improved online/offline signature schemes,'' in
  \emph{Annual International Cryptology Conference}.\hskip 1em plus 0.5em minus
  0.4em\relax Springer, 2001, pp. 355--367.

\bibitem{reyzin2002better}
L.~Reyzin and N.~Reyzin, ``Better than {BiBa}: Short one-time signatures with
  fast signing and verifying,'' in \emph{Australasian Conference on Information
  Security and Privacy}.\hskip 1em plus 0.5em minus 0.4em\relax Springer, 2002,
  pp. 144--153.

\bibitem{Aysu2016}
A.~{Aysu} and P.~{Schaumont}, ``Precomputation methods for hash-based
  signatures on energy-harvesting platforms,'' \emph{IEEE Transactions on
  Computers}, vol.~65, no.~9, pp. 2925--2931, 2016.

\bibitem{ateniese2013low}
G.~Ateniese, G.~Bianchi, A.~Capossele, and C.~Petrioli, ``Low-cost standard
  signatures in wireless sensor networks: a case for reviving pre-computation
  techniques?'' in \emph{Proceedings of NDSS 2013}, 2013.

\bibitem{elgamal1985public}
T.~ElGamal, ``A public key cryptosystem and a signature scheme based on
  discrete logarithms,'' \emph{IEEE transactions on information theory},
  vol.~31, no.~4, pp. 469--472, 1985.

\bibitem{gallagher2013digital}
P.~Gallagher, ``Digital signature standard ({DSS}),'' \emph{Federal Information
  Processing Standards Publications, volume FIPS}, pp. 186--3, 2013.

\bibitem{even1989line}
S.~Even, O.~Goldreich, and S.~Micali, ``On-line/off-line digital signatures,''
  in \emph{Conference on the Theory and Application of Cryptology}.\hskip 1em
  plus 0.5em minus 0.4em\relax Springer, 1989, pp. 263--275.

\bibitem{krawczyk2000chameleon}
H.~M. Krawczyk and T.~D. Rabin, ``Chameleon hashing and signatures,'' Aug.22
  2000, {US} Patent 6,108,783.

\bibitem{yang2020lis}
Z.~Yang, C.~Jin, Y.~Tian, J.~Lai, and J.~Zhou, ``Li{S}: Lightweight signature
  schemes for continuous message authentication in cyber-physical systems,'' in
  \emph{Proceedings of the 15th ACM Asia Conference on Computer and
  Communications Security}, 2020, pp. 719--731.

\bibitem{ParkAmortization}
{Jung Min Park}, E.~K.~P. {Chong}, and H.~J. {Siegel}, ``Efficient multicast
  packet authentication using signature amortization,'' in \emph{Proceedings
  2002 IEEE Symposium on Security and Privacy}, 2002, pp. 227--240.

\bibitem{liu2012pkc}
Y.~Liu, J.~Li, and M.~Guizani, ``Pkc based broadcast authentication using
  signature amortization for wsns,'' \emph{IEEE Transactions on Wireless
  Communications}, vol.~11, no.~6, pp. 2106--2115, 2012.

\bibitem{tartary2011authentication}
C.~Tartary, H.~Wang, and S.~Ling, ``Authentication of digital streams,''
  \emph{IEEE Transactions on Information Theory}, vol.~57, no.~9, pp.
  6285--6303, 2011.

\bibitem{yavuz2014efficient}
A.~A. Yavuz, ``An efficient real-time broadcast authentication scheme for
  command and control messages,'' \emph{IEEE Transactions on Information
  Forensics and Security}, vol.~9, no.~10, pp. 1733--1742, 2014.

\bibitem{lamport1979constructing}
L.~Lamport, ``Constructing digital signatures from a one-way function,''
  Technical Report CSL-98, SRI International Palo Alto, Tech. Rep., 1979.

\bibitem{rabin1979digitalized}
M.~O. Rabin, ``Digitalized signatures and public-key functions as intractable
  as factorization,'' Massachusetts Inst of Tech Cambridge Lab for Computer
  Science, Tech. Rep., 1979.

\bibitem{merkle1989certified}
R.~C. Merkle, ``A certified digital signature,'' in \emph{Conference on the
  Theory and Application of Cryptology}.\hskip 1em plus 0.5em minus 0.4em\relax
  Springer, 1989, pp. 218--238.

\bibitem{perrig2001biba}
A.~Perrig, ``The {BiBa} one-time signature and broadcast authentication
  protocol,'' in \emph{Proceedings of the 8th ACM conference on Computer and
  Communications Security}.\hskip 1em plus 0.5em minus 0.4em\relax ACM, 2001,
  pp. 28--37.

\end{thebibliography}


\end{document}